# Defining Tourism Domains for Semantic Annotation of Web Content

Oleksandra Panasiuk[a], Elias Kärle[a], Umutcan Simsek[a], Dieter Fensel[a]

[a] Semantic Technology Institute
Universität Innsbruck, Austria
{oleksandra.panasiuk, elias.kaerle, umutcan.simsek, dieter.fensel}@sti2.at

**Abstract**

Schema.org is an initiative by Bing, Google, Yahoo! and Yandex that publishes a vocabulary for creating structured data markup on web pages. The use of schema.org is necessary to increase the visibility of a website, making the content understandable to different automated agents (e.g. search engines, chatbots or personal assistant systems). The domain specifications are the subsets of types from the schema.org vocabulary, each associated with a set of properties. The challenge is to choose the right classes and properties for an annotation in a given domain. In this paper we address the problem of finding a subset of types and properties for complete and correct annotation of different tourism domains. The approach provides a collection of domain specifications that were built based on domain analysis and vocabulary selection.

## 1 Introduction

For the process of creating and publishing structured data on the web, it is necessary to find and use the most suitable widely acknowledged vocabulary. Since 2011 Schema.org became the de-facto standard for embedding semantic data into webpages. Bing, Google, Yahoo! and Yandex propose using the schema.org vocabulary along with the Microdata, RDFa, or JSON-LD formats to markup website content. Such a markup can be recognised by search engines and different automated agents, thus gaining access to the meaning of the sites (text, image, video, and location) and giving the possibility of returning better search results. This paper describes a methodology for defining domain specific subsets of types and properties from the schema.org vocabulary. We demonstrate our approach for the tourism domain. For our study we identified the major touristic service types and carefully analysed their online presence and existing annotations. We provided a selection of subsets from schema.org, based on Google Search Feature Gallery[1] recommendations, our methodology and results of testing and validating the structured data with the Structured Data Testing Tool[2]. As a pilot we used domains and annotations of destination marketing organisations (DMOs), such as: Tourismusverband Seefeld[3], Tourismusverband Erste Ferienregion im Zillertal[4], and Tourismusverband Mayrhofen-Hippach[5] as described in [Akbar et al., 2017]. To improve the use and to tailor schema.org for specific needs in certain application scenarios, as well as making the annotation process easier, especially for non-technical users, we propose a solution with a set of domain specifications. It gives a standard for providing correct and full annotations, increases the quality of structured data on the web, and is useful for the annotation validation and automatic generation of user interfaces (e.g. annotation editors). The remainder of this paper is structured as follows: Section 2 presents a list of work related to the approach presented in this paper. Section 3 describes the methodology for the domain definition. Section 4 presents the results of the work. Section 5 concludes the paper and outlines the ongoing and future work.

---

[1] https://developers.google.com/search/docs/guides/search-gallery

[2] https://search.google.com/structured-data/testing-tool

[3] https://www.seefeld.com

[4] https://www.best-of-zillertal.at

[5] https://www.mayrhofen.at

## 2 Related Work

In this section we give an overview of the related work. To reach our goal we build a strategy based on the analysis of the existing touristic service types and tourist preferences [Gretzel, 2004], [Grün, 2017], the principles of domain analysis, analysis of semantic vocabularies and ontologies, mapping, specification of mapping and validation methods. There is previous work on existing vocabularies for the tourism domain. The Harmonise project [Dell'Erba, et al., 2002] proposes an ontology-based mediation and harmonisation tool to allow touristic organisations exchange data while keeping their own data format. GoodRelations[6] is an ontology for annotating offerings and other aspects of e-commerce on the web and the official e-commerce model of schema.org. The idea of domain analysis from a software engineering perspective was first described in [Arango, 1989] and the approaches of the domain analysis method were explored in [Hjørland, 2002]. The adoption and evaluation of schema.org, based on the availability of data deployed on the web using a given standard and on empirical analysis, were researched in [Meuser et al., 2015]. Some further work related to our approach is connected to RDF validation methods: [Prud'hommeaux et al., 2014] describes the Shape Expression definition language to enable RDF validation through the declaration of constraints on the RDF model and [Gayo et al., 2015] shows the implementation of Shape Expressions adapted to RDF graphs.

## 3 Methodology

This section describes the methodology for defining domain specific subsets of types and properties from the schema.org vocabulary, to make the annotation process easier. The methodology is based on the three essential parts: (1-5) describes the annotation process, (6) domain specification modelling and (7-8) applies the constructed models.

1. Analysis of real world objects. This step characterises a subject area and the study of it based on the real world representation. For each selected subject one has to choose existing objects and service types, which fully describe it. For this step we analyse the touristic area and extract the existing service types relevant in tourism, such as: Hotels, Events, Food and Drink Establishments, Tourism Information Centers, Tourist Attractions, Wellness, Points of Interest, Infrastructure Services, Blogs and Articles.

---

[6] http://purl.org/goodrelations/

2. Domain analysis of the online representation. This step determines the important concepts for previously selected service types of a given subject area. We identify the types of web pages and the kind of information that is presented there. We go through the content, make the list of categories and subcategories and select the most important. With domain analysis we find out what data are suitable and important for the annotation process.

3. Domain definition and mapping to domain specific subsets from schema.org. In this step we construct the set of domains which we want to annotate based on the analysis explained above. The main challenge here is to find the best way to map them correctly, i.e. to choose the right class with properties.

4. Annotation development and deployment. To annotate content from a particular source we use the content data structure. If content is arranged with a structured data format then the annotation can be performed automatically. If not, then a manual annotation needs to be done. The automatic annotations can be developed based on a domain specific editor, wrapper, and a semantic validation mechanism and deployed by using a publishing tool.

5. Evaluation and analysis of the annotations. After the annotations are deployed, we regularly monitor their impacts on search engine results, especially rich results on Google Search. For quantitative evaluations it is good to use the Google's Search Console[7], with which it is possible to measure how much time was required to detect the annotated pages, how often the annotated pages were crawled, and how many errors were detected.

6. Domain specification modelling. This part shows how to create the common model for annotating different touristic domains. For this purpose we choose the wide use domains for a given subject area. For each domain we search for suitable and correct schema.org classes, define a selection of these properties and select the range types. This part is wholly based on the knowledge gained from the domain analysis, mapping outcomes and the validation and evaluation results. Each domain specification includes the type of domain, required and optional properties from schema.org, range and attributes as shown in Table 1. To facilitate the domain specifications the Domain Definition Interface is used. To validate them we use Specific Domain Definition Validator and Structured Data Testing Tool. [Şimşek et al., 2017]

| Domain Types | Property | Range Type | Attributes |
|---|---|---|---|
| FoodEstablishment | identifier | Text or URL | required |
| Bakery | name | Text | required |
| BarOrPub | description | Text | required |
| Brewery | address | PostalAddress | required |
| CafeOrCoffeeShop | geo | GeoCoordinates | optional |
| FastFoodRestaurant | telephone | Text | required |
| IceCreamShop | email | Text | required |
| Restaurant | url | URL | required |
| Winery | priceRange | Text | required |

---

[7] https://www.google.com/webmasters/tools/

| | openingHours Specification | OpeningHoursSpecification | required, multitype |
|---|---|---|---|
| | image | ImageObject or URL | required, multitype |
| | hasMenu | Menu or Url or Text | optional |
| | acceptsReservations | Boolean, Text, Reservation | optional |
| | servesCuisine | Text | required |
| | sameAs | URL | optional, multitype |

**Table 1.** An example of the domain specification for types of Food Establishment

7. Mapping according to domain specifications.
8. Annotation development according to domain specifications.

## 4 Results

In our approach we tackled a challenge of selecting and refining a proper subset from the schema.org vocabulary to make the annotation process easier for users. The proposed methodology describes step by step the process of creating annotation and its optimisation and improvement. This methodology can be applied to different subject areas. As output, it provides a set of domain specifications, which are also created as JSON files and can be integrated to different interfaces and systems. Domain specifications are oriented on different domains and use cases, and consist of the types and properties from schema.org, which are recommended or required for creating correct annotations. These domain specifications form a common model for annotating the object services (e.g. Hotel, Event, Restaurant, Event, Tourist Attraction, Landform, and Service) and can be used for different purposes. Based on the developed domain specifications the quality of the content will increase as it becomes well-formed and semantically consistent, and the annotation process becomes easier for non-technical users. 24 domain specifications of different touristic services are available on the semantify.it platform[8] and users can easily create their own annotations using the editor with already integrated domain specifications [Kärle et al., 2017].

## 5 Conclusion and Future Work

In this paper, we present the domain specifications to provide a better quality of structured data and to make the annotation process easier for inexperienced users. We provide the domain specifications for different domains which give the recommended standard for annotation and define the model of the structure data on the web for tourism. Those domain specifications help users to make correct annotations or to improve already existing annotations. This consequently increases the online visibility of their web resources, which will appear at the top of search engines' results for a relevant query, and makes the web content understandable to different automated agents (e.g. search engines, chatbots or personal assistant systems). The domain specifications are based on subsets of types and properties from schema.org. The domain specification files can be integrated to annotation editors and used as platforms for creating and publishing structured data. In future work we want to extend the use case and provide domain specifications for other objects, such as: hiking routes, ski tours, ski slopes, ski lifts, sport activity types, as well as other kinds of infrastructure services. The goal is also to approve and evaluate the use of the domain specifications for creating annotations and to produce higher quality annotations. In addition, we want to compare amount of time spent to annotation process with and without domain specifications.

---

[8] https://www.semantify.it